\documentclass[12pt]{article}

\begin{document}
\setcounter{page}{1}
\renewcommand{\thefootnote}{\fnsymbol{footnote}}
\pagestyle{plain} \vspace{1cm}
\begin{center}
\large{\bf Some aspects of entropic gravity in the presence of a
noncommutative Schwarzschild-deSitter black hole}

\small \vspace{1cm} {\bf S. Hamid
Mehdipour\footnote{mehdipour@liau.ac.ir}}
\\
\vspace{0.5cm} {\it Department of Physics, Lahijan Branch, Islamic
Azad University, P. O. Box 1616, Lahijan, Iran }\\

\end{center}
\vspace{1.5cm}

\begin{abstract}
We study some features of entropic force approach in the presence
of a noncommutative Schwarzschild-deSitter black hole. In this
setup, there exists a similarity between the small and large
scales. There are two finite cut-off in very short and long
distances wherein the force and energy graph stop abruptly at
those scales. We find that the existence of a deSitter core around
the origin, induced by noncommutativity, in addition to a standard
deSitter background at large scale may lead to a violation of the
equivalence principle. Finally in order to directly observe the
finite cut-off at short-scale gravity, caused by noncommutativity
quantum fluctuations, we derive an effective gravitational
constant.
\end{abstract}
\newpage

\section{\label{sec:1}Introduction}
There are many proofs signifying a profound relation between
thermodynamics and the general theory of relativity. Discovery of
black hole radiation demonstrated that the black hole behaves as a
thermal system \cite{haw}. The thermodynamic laws of black holes
suggest a meaningful connection between gravity and thermodynamics
\cite{bar}. In 1995, Jacobson derived the Einstein field equation
from the first law of thermodynamics \cite{jac}. Recently,
Padmanabhan attained the Einstein field equation by uniting the
equipartition law of energy and the holographic principle
\cite{pad1}. In addition Verlinde illustrated gravity as an
entropic force, as a result of alterations in the information
related to the locations of material bodies \cite{ver}. He
acquired an effective force acting on a test mass coming near to a
holographic screen, caused by the alteration of entropy on the
screen, which satisfies the Newton's second law for gravitational
force. Verlinde's proposal has extensively been debated in the
literature \cite{ent}.

Recently, we investigated some aspects of the entropic essence of
gravity in the presence of noncommutative Schwarzschild
\cite{meh1} and Reissner-Nordstr\"{o}m \cite{meh2} black holes by
performing the method of coordinate coherent states representing
smeared structures. This method of noncommutativity is the
so-called {\it{noncommutative geometry inspired model}} (for a
review see \cite{nic1}). The eliciting of metrics for
noncommutative geometry inspired black holes is established upon
the feasible running of the minimal observable length in general
relativity. Based on this new model of noncommutativity of
coordinates, which performs the Gaussian distribution of coherent
states, the Einstein tensor in gravity field equations remains
intact but the energy-momentum tensor takes a new form. In fact,
due to the emergence of extreme energies at short distances of a
noncommutative manifold, the effects of manifold quantum
fluctuations become visible and prohibit any measurements to find
a particle position with an accuracy more than an inherent length
scale, such as the Planck length, and this means that the concept
of locality is violated \cite{sma}. Accordingly, a point-like
particle in a noncommutative spacetime is no longer modelled by a
Dirac-delta function distribution, but will be characterized as a
smeared-like particle by a Gaussian distribution of minimal width
$\sqrt{\theta}$ {\footnote{The value of $\sqrt{\theta}$ without
appearing the extra-dimensions scenarios is a value of the order
of the Planck length, i.e. $\sqrt{\theta}\sim 10^{-33}$ cm.}},
where $\theta$ is the smallest fundamental unit of an observable
area in the noncommutative coordinates, beyond which coordinate
resolution is ambiguous.

In addition, noncommutative solutions for black holes smoothly
incorporate the deSitter core around their origin into an ordinary
metric of the black hole far away from its origin \cite{nic2}. Thus,
as an impressive outcome of this noncommutative solution, the
curvature singularity at the origin of black holes is eliminated. In
lieu of the curvature singularity, a regular deSitter vacuum state
will be formed regarding the influence of the strong quantum
fluctuations at short distances in a noncommutative manifold. This
approach descends to a usual metric at large distances where the
demeanor of the minimal length is insignificant, while providing new
physics which appears at short distances.

On the other hand, since in the noncommutative geometry inspired
solutions the existence of a deSitter core in the centre of black
holes prohibits their collapse into a singular case, it can shed
more light on the issue of the quantum stability of the deSitter
space and accordingly, the rate of the Planck size black holes
production on the inflationary background of the universe
\cite{nic3}. In inflationary epochs, the universe is well
delineated by the deSitter geometry. The accelerating phase in the
inflationary era of the universe was initiated as a plan to find a
solution to the problems in the standard big-bang theory
\cite{guth}. The most significant observational development in
cosmology is the conclusion of the cosmic accelerating expansion
of the universe which was first declared in 1998, based on
Supernova data and cosmic microwave background observations
\cite{perl}. These data mention the appearance of some background
form of the energy with a negative pressure. It is possible to
illustrate this energy through a positive cosmological constant
and quintessence fields. This means that, when taking objects like
black holes into account one can presume the emergence of an
effective, positive cosmological constant. In this paper, we would
like to extend our previous work \cite{meh1} to a deSitter
background caused by a cosmological constant. It is clear that, in
light of the facts mentioned above, the interest in considering
the noncommutative Schwarzschild-deSitter black hole (NC SdS BH)
becomes natural.

\section{\label{sec:2}Noncommutative Schwarzschild-deSitter Metric}
The NC SdS BH solution obtained by Mann and Nicolini \cite{nic3}
is given by the following metric {\footnote{We use units with the
following definitions: $\hbar = c = k_B = 1$.}},
\begin{equation}
\label{mat:1}ds^2=-N(r)dt^2+ N^{-1}(r)dr^2+r^2 d\Omega^2,
\end{equation}
where
\begin{equation}
\label{mat:2}N(r)=1-\frac{2GM_\theta}{r}-\frac{\Lambda}{3}r^2.
\end{equation}
In the above, the cosmological constant is $\Lambda=3/l^2$, and $l$
is the cosmological length associated with the $\Lambda$. The
smeared mass distribution $M_{\theta}$ is found to lead to the
result
\begin{equation}
\label{mat:3}M_{\theta}=M\left[{\cal{E}}\left(\frac{r}{2\sqrt{\theta}}\right)
-\frac{ r}{\sqrt{\pi\theta}}e^{-\frac{r^2}{4\theta}}\right].
\end{equation}
For the commutative case, $r/\sqrt{\theta}\rightarrow\infty$, the
smeared-like mass descends to the point-like mass, i.e.
$M_{\theta}\rightarrow M$, and one recovers the standard SdS
metric. In fact, this is the regime where noncommutative
fluctuations are insignificant and the background geometry may
well characterized through a smooth differential manifold. However
for the noncommutative case, $r\rightarrow\sqrt{\theta}$, the NC
SdS metric deviates crucially from the standard one and provides
novel physics at short distance regime.

\begin{figure}[htp]
\begin{center}
\includegraphics{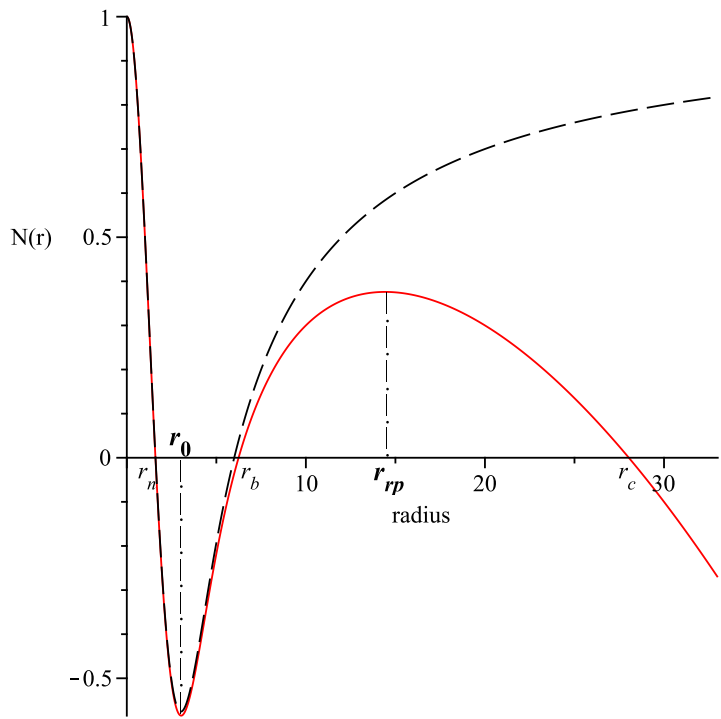}
\end{center}
\vspace{9.3 cm} \caption{\scriptsize {The function $N(r)$ versus
the radius, $r/\sqrt{\theta}$ for $M=3.0\sqrt{\theta}/G$. The
solid line corresponds to the NC SdS BH for
$\Lambda/3=10^{-3}/\theta$, and the dashed line corresponds to the
NC Schwarzschild BH ($\Lambda=0$).}} \label{fig:1}
\end{figure}
For further details, we draw the temporal component of the metric
(\ref{mat:1}), $N(r)$, as a function of $r/\sqrt{\theta}$ (see
Fig.~\ref{fig:1}), for $M=3.0\sqrt{\theta}/G$, in two cases: a)
deSitter background for $\Lambda/3=10^{-3}/\theta$ and; b)
asymptotically flat space. The solid line presented in
Fig.~\ref{fig:1} shows the possibility of having three distinct
horizons at a typical case when the mass of the black hole is
larger than the minimal nonzero mass $M_0$ but smaller than the
Nariai mass $M_N$. Note that, for $M>M_N$, there is no timelike
Killing vector, and for $M<M_0$, there is no black hole
\cite{nic3}. Intersections on the radius axis lead to radii of the
event horizons. For the case a) we have three horizons, an inner
(or noncommutative) $r_n$ and an outer black hole horizon $r_b$
and a cosmological event horizon $r_c$. But, for the case b) there
are only two horizons, a noncommutative and a black hole horizon
(see the dashed line). In comparison with the commutative one, for
$M \gg M_0$, the noncommutative horizon tends to zero, while the
black hole horizon reaches the Schwarzschild value, $r_b
\rightarrow 2M$.

According to the NC approach \cite{nic3}, due to the presence of
strong quantum fluctuations in a noncommutative manifold, a
regular behaviour at the origin proves natural. Since there exists
a outward push caused by noncommutativity quantum fluctuations,
the metric (\ref{mat:1}) is well characterized close to the origin
by a deSitter geometry, this can be illustrated by a quantum
pressure that is related to the cosmological constant in deSitter
universe. As a result, an effective cosmological constant
corresponding to the deSitter type solution, i.e.
$N(r)\approx1-\Lambda_{\textrm{eff}}\,r^2/3$, can be generated by
using the asymptotic form of the metric (\ref{mat:1}) at short
distances as follows:
\begin{equation}
\label{mat:4}\Lambda_{\textrm{eff}}=\Lambda+\frac{MG}{\sqrt{\pi\theta^3}},
\end{equation}
which is used to illustrate the accelerating expansion of the
universe. As mentioned above, the regularisation is due to a local
deSitter spacetime caused by the standard deSitter background,
plus the noncommutative fluctuations.

The authors of Ref.~\cite{sds} investigated the entropic
formulation in the presence of the SdS BH as a model of multiple
holographic screens. Due to the vanishing of the Unruh-Verlinde
temperature at the Bousso-Hawking reference point \cite{bou}, they
considered two regions separated by zero temperature barrier as
thermodynamically isolated systems in a static geometry setup, and
finally applied independently the Verlinde's entropic formalism to
each region. In this work, we utilize the Bousso-Hawking reference
point to observe the temperature on the holographic screens.
However, we only consider the internal region restricted by
surfaces at $r=r_0$ and $r=r_{rp}$, i.e. the pattern of the metric
for $r_0\leq r\leq r_{rp}$, where $r_0$ and $r_{rp}$ are the
minimal nonzero radius and the reference point radius,
respectively (see Fig.~\ref{fig:1}). We will explain this issue in
the next section.

\section{\label{sec:3}Verlinde's Entropic Formalism}
In the curved spacetime, to look for a time-like Killing vector
$\xi_\alpha$ of the NC SdS BH which is asymptotically de Sitter
space, we consider a normalization constant $\sigma$ for the
$\xi_\alpha$, where
\begin{equation}
\label{mat:5}\xi_\alpha=\sigma(\partial_0)_\alpha.
\end{equation}
In the asymptotically flat spacetime, the standard Killing vector
normalization, i.e. $\sigma=1$, is recovered. In order to prevent
a complication in taking the normalization of Killing vector, we
use the normalization proposed by Bousso and Hawing \cite{bou} so
that the norm of the Killing vector is one at the area where the
gravitational attraction and the cosmological repulsion cancel
each other and thus the force vanishes. Since SdS space is not
asymptotically flat, they placed a reference point in the radial
direction such that it can fulfill a role of a point at infinity
in an asymptotically flat spacetime. Moreover, the temperature at
this point vanishes, and any thermal exchanges cannot transpire
through the reference point. Therefore, a thermally insulating
wall is made at that region.

\begin{figure}[htp]
\begin{center}
\includegraphics{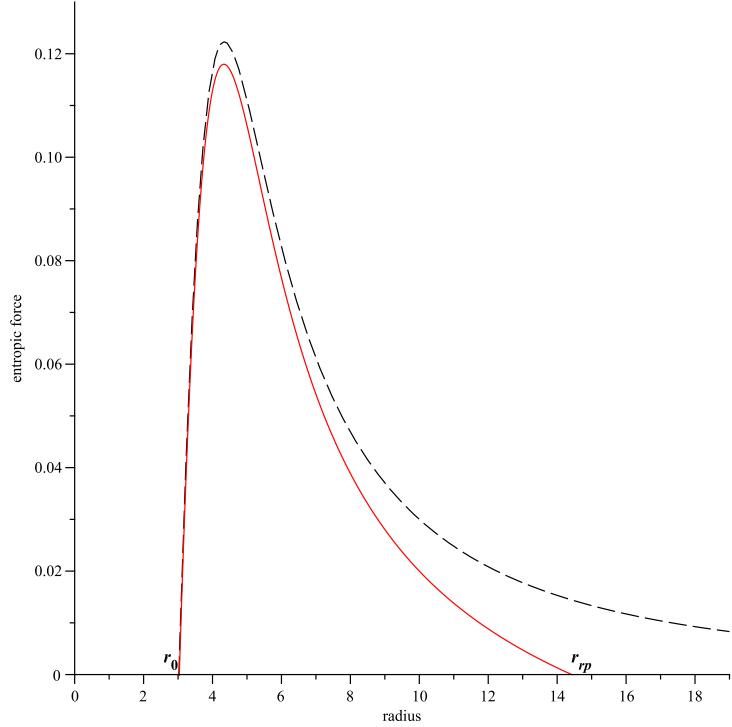}
\end{center}
\vspace{7.3 cm} \caption{\scriptsize {The entropic force $F$
versus the radius, $r/\sqrt{\theta}$. We have set
$M=3.0\sqrt{\theta}/G$. The solid line represents the entropic
force in a deSitter background for $\Lambda/3=10^{-3}/\theta$. The
dashed line represents the entropic force of the NC Schwarzschild
BH in asymptotically flat space. }} \label{fig:2}
\end{figure}

\begin{figure}[htp]
\begin{center}
\includegraphics{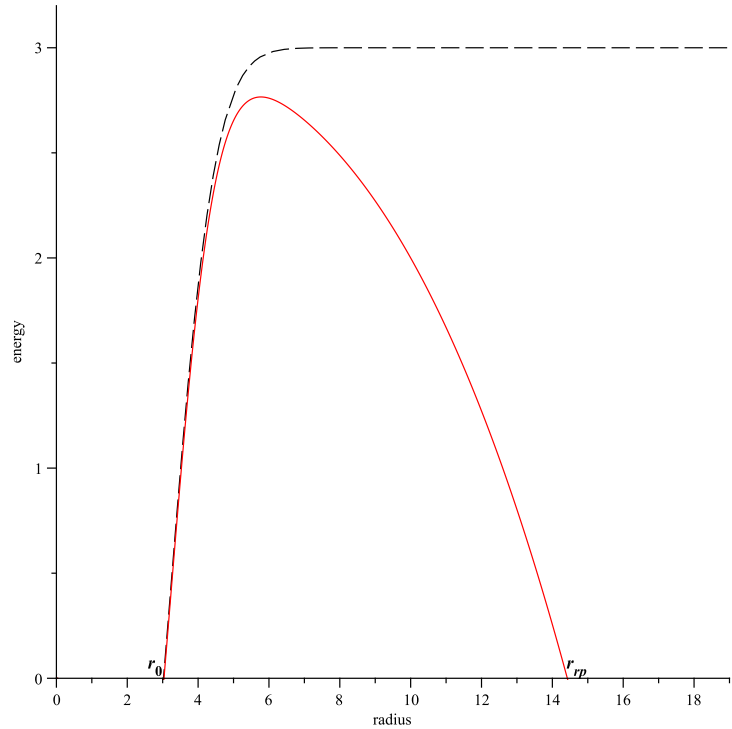}
\end{center}
\vspace{8.5 cm} \caption{\scriptsize {The energy,
$E/\sqrt{\theta}$, versus the radius, $r/\sqrt{\theta}$. We have
set $M=3.0\sqrt{\theta}/G$. The solid line corresponds to the NC
SdS BH for $\Lambda/3=10^{-3}/\theta$, and the dashed line
corresponds to the NC Schwarzschild BH ($\Lambda=0$).}}
\label{fig:3}
\end{figure}

Since we need to distinguish holographic screens $\Omega$ at
surfaces of constant redshift and in order to clarify a foliation
of space, the generalized form of the Newtonian potential $\phi$
and the acceleration $a^\alpha$ in the general relativistic
framework can be written as
\begin{equation}
\label{mat:6}\phi=\frac{1}{2}\log\left(-g^{\alpha\beta}\xi_\alpha\xi_\beta\right),
\end{equation}
\begin{equation}
\label{mat:7}a^\alpha=-g^{\alpha\beta}\nabla_\beta\phi,
\end{equation}
Using the Killing equations
$\partial_\alpha\xi_\beta+\partial_\beta\xi_\alpha=2\Gamma^\gamma_{\alpha\beta}\xi_\gamma$,
here $\alpha, \beta, \gamma$ run from $0$ to $3$, with the
condition of static spherically symmetric
$\partial_0\xi_\alpha=\partial_3\xi_\alpha=0$, and also the
normalization mentioned above, the gravitational potential for the
NC SdS BH is found to have the form
\begin{equation}
\label{mat:8}\phi=\frac{1}{2}\log\left(\sigma^2N(r)\right),
\end{equation}
where $e^\phi$ is the redshift factor and is equal to one at the
the Bousso-Hawking reference point. The Unruh-Verlinde temperature
on the screen can be written in the form
\begin{equation}
\label{mat:10}T=-\frac{1}{2\pi}e^\phi n^\alpha
a_\alpha=\frac{e^\phi}{2\pi}\sqrt{g^{\alpha\beta}\nabla_\alpha\phi\nabla_\beta\phi},
\end{equation}
where $n^\alpha$ is a unit vector which is defined as
\begin{equation}
\label{mat:11}n^\alpha=\frac{\nabla^\alpha\phi}{\sqrt{g^{\alpha\beta}\nabla_\alpha\phi\nabla_\beta\phi}}.
\end{equation}
The unit vector $n^\alpha$ is normal to the holographic screen and
to the $\xi_\alpha$. The existence of the expression $e^\phi$ in
Eq.~(\ref{mat:10}) is due to the fact that the temperature is
measured with respect to the reference point. In our case, the
reference point is placed at the region between the black hole
horizon and the cosmological event horizon wherein the force is
zero. The Unruh-Verlinde temperature for the NC SdS BH has the
form
\begin{equation}
\label{mat:12}T=\sigma\frac{N'(r)}{4\pi}=\frac{\sigma}{2\pi}\left(\frac{GM_{\theta}}{
r^2}-rh(r)\right),
\end{equation}
where the prime abbreviates $d/dr$, and
\begin{equation}
\label{mat:13}h(r)=\frac{GM}{2\sqrt{\pi\theta^3}}e^{-\frac{r^2}{4\theta}}+\frac{\Lambda}{3}.
\end{equation}
The energy on the holographic screen $\Omega$, according to the
equipartition law of energy, is immediately written as
\begin{equation}
\label{mat:14}E=\frac{1}{4\pi}\int_\Omega e^\phi\nabla\phi dA=2\pi
r^2T,
\end{equation}
where $A$ is the area of the screen. For the energy on the NC SdS
screen, one can find
\begin{equation}
\label{mat:15}E=\sigma\left(GM_{\theta}-r^3h(r)\right).
\end{equation}
The entropic force is therefore given by
\begin{equation}
\label{mat:16}F_\alpha=T\nabla_\alpha S,
\end{equation}
where $\nabla_\alpha S=-2\pi m n_\alpha$, is the change in entropy
for the test mass at fixed position nearby the screen. Ultimately,
the entropic force in the presence of the NC SdS BH becomes
\begin{equation}
\label{mat:17}F=\sqrt{g^{\alpha\beta}F_\alpha
F_\beta}=\sigma\left( \frac{GM_{\theta}m}{r^2}-mrh(r)\right).
\end{equation}
The numerical results of the entropic force and the energy versus
the radius for two situations, a deSitter background and the
asymptotically flat spacetime, are shown in Figs.~\ref{fig:2} and
\ref{fig:3}, respectively. We notice that all of
Figs.~\ref{fig:1}, \ref{fig:2} and \ref{fig:3} emphasize the
region between two boundaries at radii $r_0$ and $r_{rp}$ such
that the temperature, force and energy vanish on these boundaries.
As can be seen from last two figures, the force and energy graph
cut off abruptly at some finite $r$ at both small and large
scales.

About the small scales, as we have already mentioned in
Refs.~\cite{meh1,meh2}, due to a negative quantum pressure induced
by the coordinate noncommutativity at small scales, the case of
$r<r_0$ leads to some out of the ordinary dynamical features like
negative entropic force, i.e. gravitational repulsive force, and
negative energy. According to the original work proposed by
Nicolini {\it et al} (see the first reference of \cite{nic2}), if
we choose that the original mass is less than the minimal mass
$M_0$, or the screen radius to be less than the radius of the
smallest holographic surface at the Planckian regime, there cannot
be a black hole and no temperature can be defined, thus the
ultimate zero temperature configuration can be observed a black
hole remnant at the region where the Hawking emission stops
abruptly; as a result, in agreement to Verlinde's entropic
formalism, the behavior of the entropic force is similar to the
temperature. Moreover if $r<r_0$, we encounter with the unusual
case of $E<0$ which is nonphysical, so the existence of finite
cut-off at small scales is credible and one can make the
requirement that $E\geq 0$.

There is also a same reason for the large scales. In fact, the
pattern of the metric for very short distances has a similarity to
the pattern of the metric for very long distances. Here we imply
that the standard deSitter background at large scales may prevent
a measurement to find the particle position beyond the reference
point radius, a maximal observable length, in exactly the same way
that the existence of a deSitter core in the centre of the black
hole yields a outward push to prevent its collapse into a singular
case. Therefore, it is impossible to set up a measurement to find
more accurate particle position than $r_0$. This means that if $r$
is too small or large, as the test mass $m$ comes close to the
screen, the lessening in screen entropy will produce a repulsive
force. Thus, it is not necessary to consider the total system and
one can ignore the patterns of the metric for $r<r_0$ and
$r>r_{rp}$. Consequently, we apply the circumstance that the
screen radius is bigger than the radius of the smallest
holographic surface but is smaller than the radius of the
Bousso-Hawking reference point.

As mentioned above, in agreement to our previous works for small
scales \cite{meh1,meh2}, Figs.~\ref{fig:2} and \ref{fig:3} show
that the entropic force and energy on the holographic screens with
radii $r_0$ and $r_{rp}$ are zero. This is an important result due
to the fact that $r_0$ and $r_{rp}$ are, respectively, radii of
smallest and largest holographic screens, then they cannot be
probed through the test mass that is located on a very short or
long distance from the source. Accordingly, the conventional
formulation of gravity is contravened in both small and large
scales when the screen radius comes near the $r_0$ or $r_{rp}$. In
other words, the test mass cannot recognize any gravitational
field in two situations: i) when it is located at a minimal
distance from the source mass; ii) when it is located at a maximal
distance from the source mass. The circumstance adopted by the
first situation manifestly contravenes the entity of the
exclusively gravitational interaction for an inert remnant of the
black hole. Black hole residues as crucial physical entities are
profoundly confirmed in the quantum gravity literature when
quantum gravitational fluctuations are exposed. As an example,
when generalized uncertainty principle is taken into
consideration, the total evaporation of black holes is banned and
there would be massive but inert residues including the
exclusively gravitational interactions \cite{adl}. Our approach
proves that the black hole remnants are totally inert with no
gravitational interactions. This enables one to locally mark a
difference between a uniform acceleration and a gravitational
field. When one reaches the smallest fundamental unit of a
holographic screen with radius $r_0$ one conflict with the
equivalence principle (EP) of general relativity because there is
now an essential distinction between the gravitational and
inertial mass. In fact, contrary to the inertial mass, the
gravitational mass in the black hole remnant possesses no
gravitational field which is recognized to be zero. Therefore,
there may be a violation of the EP at small scales owing to the
existence of a deSitter core around the origin. In a similar
manner, one can imagine that an evident violation of the EP may
occur owing to the existence of a standard deSitter background at
large scales. This means that it may be possible to observe a
distinction between the gravitational and inertial mass in a
locally frame of reference at a cosmic size. Such an argument,
though perhaps too far to be perceivable directly, could in
principle leaves a trace of a unified theory at all length scales.

\begin{figure}[htp]
\begin{center}
\includegraphics{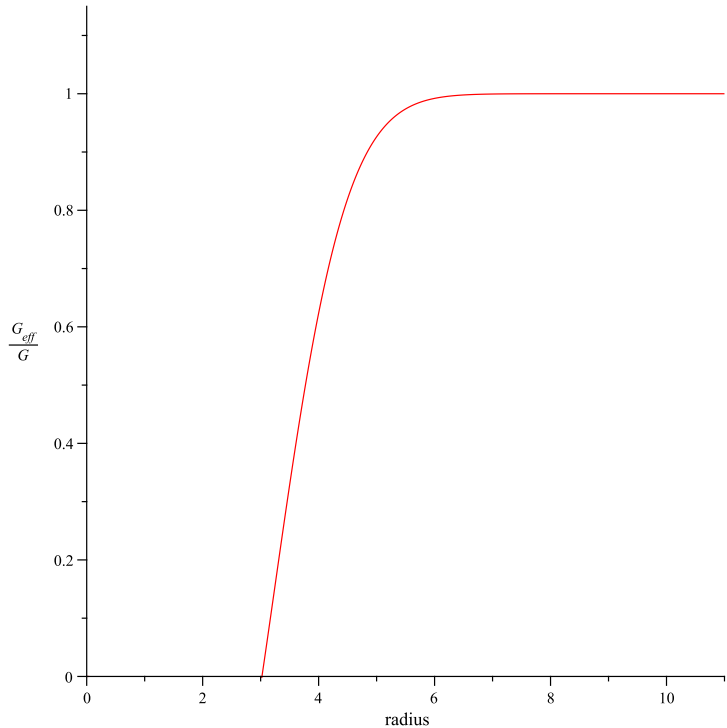}
\end{center}
\vspace{7.3 cm} \caption{\scriptsize {The function
$G_{\textrm{eff}}/G$ versus the radius, $r/\sqrt{\theta}$. The
fraction of gravitational constant deviates strongly from the
unity at small scales.}} \label{fig:4}
\end{figure}

Let us now return to Eq.~(\ref{mat:17}) and consider it as
\begin{equation}
\label{mat:18}F=\sigma\left(\frac{G_{\textrm{eff}}Mm}{r^2}-\frac{m\Lambda}{3}r\right).
\end{equation}
In the same manner, energy becomes
\begin{equation}
\label{mat:19}E=\sigma\left(G_{\textrm{eff}}M-\frac{\Lambda}{3}r^3\right),
\end{equation}
where $G_{\textrm{eff}}$ is defined as an effective gravitational
constant which is written as
\begin{equation}
\label{mat:20}G_{\textrm{eff}}=G\left[{\cal{E}}\left(\frac{r}{2\sqrt{\theta}}\right)
-\frac{r}{\sqrt{\pi\theta}}e^{-\frac{r^2}{4\theta}}\left(1+\frac{r^2}{2\theta}\right)\right].
\end{equation}
From the result (\ref{mat:20}) we can observe that the effective
gravitational constant incorporates effects of the
noncommutativity of coordinates and depends on the
$r/\sqrt{\theta}$\, so that in the commutative case,
$r/\sqrt{\theta}\rightarrow \infty$, we have the usual
gravitational constant, i.e. $G_{\textrm{eff}}\rightarrow G$.
Thus, we see that the noncommutative geometry inspired model can
predict an effective gravitational constant as well. The plot
presented in Fig.~\ref{fig:4} exhibits the numerical results of
the function $G_{\textrm{eff}}/G$ versus the radius,
$r/\sqrt{\theta}$.

As this figure shows, the lowering of the $G_{\textrm{eff}}$ with
respect to the $G$ near the origin is clear. The appearance of a
lower finite cut-off at short-scale gravity imposes a bound on any
measurements to find a particle position in noncommutative
geometry.

\section{\label{sec:4}Summary}
In summary, we have applied the noncommutative geometry inspired
model to include the microscopic structure of spacetime in the
entropic view of gravity by reason of significant coupling between
the issue of entropy with the quantum spacetime structures. The
entropic force in the presence of NC SdS BHs by considering the
effect of smearing of the particle mass as a Gaussian distribution
is derived. In this approach, the force and energy graph cut off
abruptly at some finite screen radii on both small and large
scales. This allows us to impose bounds on the scale of NC SdS
BHs. We have shown that when one combines entropic gravity with
noncommutative geometry there can be a violation of the EP for
both small and large scales which signals a failure of current
physical ideas. In the end, an effective gravitational constant
induced by noncommutativity parameter is derived by applying
noncommutative effects in Verlinde's formalism of gravity.

\end{document}